\documentclass{article}
\newcommand{\mcl}[1]{\mathcal{#1}}

\newcommand{\fant}[1]{\phantom{#1}}
\newcommand{\be}{\begin{equation}}
\newcommand{\ee}{\end{equation}}

\newcommand{\wdg}{\wedge}
\newcommand{\ot}{\otimes}

\begin{document}
\title{Cylindrically Symmetric-Static  Brans-Dicke-Maxwell
Solutions}
\author{Ahmet Baykal\footnote{{Department of Physics, Bogazici University,
 Bebek, Istanbul, Turkey; e-mail: baykala@boun.edu.tr}}
\  and \"{O}zg\"{u}r Delice\footnote{Department of Physics,
Bogazici University,  Bebek, Istanbul, Turkey; e-mail:
odelice@boun.edu.tr}}
\maketitle
\begin{abstract}
We present static cylindrically symmetric  electrovac solutions in
the framework of the Brans-Dicke theory and show that our solution
yields some of the well-known solutions for special  values of  the
parameters of the resulting metric functions.
\end{abstract}


\section{Introduction}
Among various versions of the scalar-tensor theories of gravity
the Brans-Dicke theory can be considered as the prominent one.
Although its original motivations are rooted in the attempts to incorporate the Machian
view of inertia into the general theory of relativity,
it has been  applied to issues apart from its
original intent, ranging from inflation schemes in cosmology
to quantum gravity \cite{brans}. The inclusion of scalar fields into general relativity
is also favored, for example, by the low energy limit of string theories in the guise
of  a dilaton field.
Similarly, scalar fields also appear as by-products of dimensional reduction procedure
in Kaluza-Klein theories. Thus, it is worth to investigate the implications of such
scalar-tensor theories for various models of space-time in comparison to general theory
of relativity.

In this paper, we will obtain static cylindrically symmetric
 electro-vacuum solutions in the framework of the Brans-Dicke
 scalar tensor theory \cite{brans-dicke} and compare the solutions with corresponding Einstein-Maxwell
 solutions. Cylindrically symmetric, static Einstein-Maxwell electro-vacuum
 solutions are well known \cite{SKMHH}. Bonnor presented general solutions with axial or
longitudinal magnetic fields \cite{Bonnor} and Raychaudhuri
presented a solution with radial electric field \cite{Raychaudhuri}.
These solutions have been also discussed  by L. Witten
 by using Rainich conditions in an ``already unified theory" \cite{Witten,Rainich}. All
these solutions yield the same static vacuum solution -the
Levi-Civita solution- when the parameters appearing in the metric
related with electric or magnetic fields vanish \cite{levicivita}. All the
solutions are singular at the axis of symmetry except the so-called
Bonnor-Melvin  magnetic universe {\cite{BonnorMelvin}}
having a uniform magnetic field along the $z$ axis. When magnetic
field vanishes this solution reduces to Minkowski space-time. To the best of our knowledge,
the only solution of this form in the framework of
Brans-Dicke theory is given in the work of  Banerjee \cite{Banerjee}, where
only a solution with a radial electric field is discussed.
In the present  work, we will consider other possible  field configurations of a non-null Maxwell field,
where there is an axial or an angular magnetic field. An
analysis  similar to our work has been made  in cylindrical symmetric model space-time for a  dilaton
scalar field coupled nonlinearly to Electromagnetic field tensor
\cite{Dilaton}. However, for the solutions therein, unlike our solutions,
when scalar field vanishes the electromagnetic field
vanishes as well, hence there is no limit such that the scalar field
vanishes and the solution reduces to Einstein-Maxwell solutions.

The paper is organized as follows. In the next section we present
our notation and the field equations. In the third section  we
present the solutions corresponding to a longitudinal,  axial
magnetic field and  a radial electric field separately.

\section{Field Equations}

 The field equations for the scalar-tensor theory of
Brans-Dicke minimally coupled  to Max\-well field can be obtained from the
Lagrangian density written in terms of exterior forms as
\be\label{lag-density}
\mcl{L}[\phi,g, A]
=
 \frac{1}{2}\phi
\Omega_{\mu\nu}\wedge*(\theta^{\mu}\wedge\theta^{\nu})
-
\,\frac{\omega}{2}\,d\phi\wedge*d\phi
-
\frac{1}{2}\,F\wedge*F,
\ee
where $\theta^\mu$ ( $ \mu=0,1,2,3$) are
the set of coframe basis one forms. As in the standard general theory of
relativity, $\Omega_{\mu\nu}$ is curvature 2-form  derived from
the torsion free metric compatible connection one form
$\omega_{\mu\nu}$. $*$ is the linear Hodge dual operator through
which matter fields couple to the metric in the coframe
formulation. $\phi$ is the Brans-Dicke scalar field which replaces
the Newtonian coupling constant $1/G$.
 In this investigation the metric is coupled only to the Faraday 2-form $F$
derived from the 4-potential $A$. The numerical factor $\omega$
contained in the dynamical term for $\phi$ is the so-called
Brans-Dicke parameter. The notation and convention in what follows
is adopted from \cite{thirring}.

The metric equations are obtained from the Lagrangian density
(\ref{lag-density})  by a suitable variational procedure with respect to the coframe fields $\theta^{\alpha}$.
This imposes the condition that the metric compatible connection one form $\omega_{\mu\nu}=-\omega_{\nu\mu}$
to be torsion-free, for example, by introducing appropriate
Lagrange multiplier 2-forms.
The field equations for  $\phi$ and $F=dA$ also follow from the variation of the
(\ref{lag-density}) with respect to $\phi$ and $A$.
The variation $\delta\mcl{L}/\delta \theta^\alpha=0$ yields the metric equations
\be\label{metric-eqn}
\phi *G^\alpha
=
D*(d\phi\wedge\theta^\alpha)
+
\,\frac{\omega}{\phi}\,*T^\alpha[\phi]
+
*T^\alpha[F]
\ee
where $G^{\alpha}=R^\alpha-1/2\theta^\alpha R$ is the Einstein one form, and
\begin{eqnarray}
*T_{\alpha}[\phi]
&=&
\frac{\omega}{2\phi}\left\{i_{\alpha}(d\phi)*d\phi+(d\phi)\wdg i_{\alpha}*d\phi\right\}
\\
{*}T_{\alpha}[F]
&=&
1/2\left\{i_{\alpha}F\wdg*F-F\wdg i_{\alpha}*F\right\},
\end{eqnarray}
are the Hodge duals of the Energy-momentum 1-forms
$T^{\alpha}=T^{\alpha}_{\fant{\alpha}\beta}\theta^{\beta}$ of the
fields $\phi$ and $F$ respectively. $i_{\alpha}=i_{e_\alpha}$ is
the contraction operator
($i_{\alpha}\theta^\beta=\delta^\beta_\alpha$) and $D$ is the
covariant exterior derivative acting on tensor valued forms.
Unlike the Maxwell field, which only couples to the metric
components, the scalar field $\phi$ couples to the derivatives of
the metric field so that one has the term
\be
D*(d\phi\wdg
\theta^{\alpha}) = d*(d\phi\wdg
\theta^{\alpha})+\omega^{\alpha}_{\fant{\alpha}\beta}\wdg*(d\phi\wdg
\theta^{\beta})
\ee
on the right-hand-side of (\ref{metric-eqn}).
This term is essential for the the conservation of matter,
$D*T^{\alpha}[F]=0$, assuming that the second Bianchi identity
holds \cite{brans-dicke}. The trace of (\ref{metric-eqn}) together
with $\delta\mcl{L}/\delta\phi=0$ yields the equation for $\phi$
\be
\label{scalar-eqn} (2\omega+3)d*d\phi=* T=0,
\ee
since the trace of the energy momentum tensor of the Maxwell field vanishes.
Added to these,
one has the Maxwell's equations \be\label{faraday-eqn}
d*F=0,\qquad dF=0 \ee which hold outside the sources. We solve
the field equations by starting with the cylindrically symmetric
static ansatzs   of the canonical form \cite{SKMHH}
\be
g=\eta_{\mu\nu}\theta^{\mu}\otimes \theta^{\nu}, \qquad
\eta^{\mu\nu}=\eta_{\mu\nu}=diag(-,+,+,+),
\ee
with  the coframe fields are given by
\be \theta^0=e^{K-U}dt,\quad
\theta^1=e^{K-U}dr,\quad \theta^{2}=e^U dz,\quad \theta^{3}=e^{-U}
W d\phi. \label{metric1}
\ee
The cylindrically symmetric line element has three commuting Killing fields so
that one has three ignorable coordinates and all the metric
functions $K, U, W$ depend on a single coordinate which is denoted
by $r$. This also dictates that  the Brans-Dicke scalar $\phi$
and the components of the 2-form $F$  should also be the functions of this
coordinate only. For a static spacetime  $dt\ot d\phi$  term of
the metric  is zero. The coordinates $t,r,z,\varphi$ are the
time-like, the radial, the axial and the angular coordinates of
the cylindrical symmetric static metric with  ranges
$-\infty<t,z,<\infty, 0\le r \le \infty$ and $0 < \varphi \le
2\pi$ respectively.

It is  also possible to consider an alternative form of the static
cylindrical metric by applying a complex substitution
$z\rightarrow it$, $t\rightarrow iz$ which yields the coframe
forms as: \be \tilde{\theta}^0=e^{U}dt,\quad
\tilde{\theta}^1=e^{K-U}dr,\quad \tilde{\theta}^{2}=e^{K-U}
dz,\quad \tilde{\theta}^{3}=e^{-U} W d\phi. \label{metric2} \ee
This form of the metric is useful \cite{SKMHH, maccallum} when we
consider a radial electric field.

With the ansatzs (\ref{metric1}) for the metric field,  the
field equations (\ref{metric-eqn}), (\ref{scalar-eqn}) become
\begin{eqnarray}
&& -\frac{W''}{W}+K'\frac{W'}{W}-U'^2 =
\frac{\omega}{2}\left(\frac{\phi'}{\phi}\right)^2 -\left\{
(K'-U')\frac{\phi'}{\phi}\right\}
+
\left\{\frac{\phi^{''}}{\phi}+\frac{W'}{W}
\frac{\phi'}{\phi}\right\}
\nonumber\\
&&\phantom{AAAAAAAAAAAAAAAAAAAA}+\frac{1}{\phi}T_{00}[F]e^{2(K-U)},\label{G11}
\end{eqnarray}
\be\label{G22}
K'\frac{W'}{W}-U'^2
=
\frac{\omega}{2}\left(\frac{\phi'}{\phi}\right)^2-\left\{(K'-U')+\frac{W'}{W}\right\}\frac{\phi'}{\phi}
+\frac{1}{\phi}T_{11}[F]e^{2(K-U)},
\ee
\begin{eqnarray}
&&\frac{W''}{W}-2U''-2U'\frac{W'}{W}+K''+U'^2=-\frac{\omega}{2\phi^2}\phi'^2-\frac{1}{\phi}
\left\{\phi^{''}+\phi'\frac{W'}{W}\right\}\nonumber\\
&&\phantom{AAAAAAAAAAAAAAAAAAA} +U'\frac{\phi'}{\phi}
+\frac{1}{\phi}T_{22}[F]e^{2(K-U)},\label{G44}
\end{eqnarray}
\be\label{G33} K''+U'^2 = -
\frac{\omega}{2}\left(\frac{\phi'}{\phi}\right)^2 -
U'\frac{\phi'}{\phi}-\frac{\phi''}{\phi} +
\frac{1}{\phi}T_{33}[F]e^{2(K-U)}
\ee
\be\label{phieq}
\phi''+\phi'\frac{W'}{W}=0,
\ee
For the metric ansatz (\ref{metric2}), the only change in
the field equations above is that one needs the exchange $T_{00}\leftrightarrow T_{22}$.
Since we will consider various configurations of the field $F$, explicit form of the Maxwells's equations
will be considered below.

\section{Solutions }
The isometries of the above metric ansatzs restrict the components of the
 Faraday tensor that couples to the metric.
In effect, the corresponding four currents $J$ have vanishing
wedge products with the Killing one-forms of the Killing vector fields
$\partial_\phi$ and $\partial_z$. That is, one has  $dz\wdg d\phi \wdg J=0$.
In the light of these observations,
one can consider $F\sim \theta^{1}\wdg \theta^{2}$ or $F\sim\theta^{1}\wdg \theta^{3}$. In addition,
corresponding to  a charge
distribution  along the symmetry axis, $F$ assumes the form $F\sim
\theta^{0}\wdg\theta^{1}$. However, although  linear combinations
of these fields also satisfy the homogeneous Maxwell's equations, not all of such linear
combinations are compatible with the metric ansatzs  to begin with.
For all the cases, the components of Faraday 2-forms are functions of the radial coordinate.

\subsection{A Magnetic field along the $z$ direction.}
 First, we  consider a magnetic field along $z$-direction,  which
 is due to  current  sources circulating about the $z$-axis. The corresponding
 closed  Faraday 2-form is of the form $F=f(r)\theta^{1}\wdg \theta^{2}$. For this configuration
 of the fields,
  the metric ansatz (\ref{metric1}) is considered. Maxwell
 equation $d*F=0$ is satisfied for $f(r)=\alpha
\frac{e^{2U-K}}{W}$ with $\alpha$ a constant. The corresponding
energy momentum tensor field has the components
\begin{equation}\label{en-mom-F1}
T_{00}[F]=T_{11}[F]=T_{22}[F]=-T_{33}[F]=\frac{1}{2}f^2.
\end{equation}
The structure of the coupled ordinary differential equations for all the fields
can be reduced to an equation for a single metric function from which all the others
can be computed. Explicitly,
from (\ref{phieq}), (\ref{G11})-(\ref{G22}) and
(\ref{G22})+(\ref{G33}) one finds
\begin{eqnarray}
\phi'=\frac{a}{W},\quad W'=\frac{b}{\phi}, \quad W\phi=r,  \quad
K'=\frac{q}{W\phi}, \label{intermediate1}
\end{eqnarray}
($a,b, q$ are integration constants)
and in addition, from (\ref{G44})+(\ref{G33}) it follows that
\begin{equation}
\frac{\phi''}{\phi}+\frac{W''}{W}-2U''-2U'\frac{W'}{W}+2U'^2+2K''+\frac{w
\phi'^2}{\phi^2}=0.\label{intermediate2}
\end{equation}
(\ref{intermediate1}), (\ref{intermediate2})  reduce to
 a differential equation, for example, for the metric function $U$. Therefore, one can
determine $U$ and integrating back one can find all the  field functions as
\begin{eqnarray}
&& U=\frac{1}{2}\left(k-1+p \right)\ln{r}-\ln{(1+c^2r^p)},\\
&&W=W_0r^k,\quad \phi=W_0^{-1} r^{1-k},\quad K=q \ln{r},
\end{eqnarray}
where the integration constants $p, q, k$ and the Brans-Dicke parameter $\omega$ are related by
\be
p^2=1+4q+(2-3k)k+2\omega[(2-k)k-1].
\ee
These yield the metric as
\begin{eqnarray}
g
&=&
(1+c^2r^p)^2
\left\{
{r^{2q+1-p-k}}(dr\ot dr-dt\ot dt)+r^{1-p+k}W_0^2 d\phi\ot d\phi
\right\}
\nonumber\\
&&
+
\frac{r^{p+k-1}}{(1+c^2r^p)^2}dz\ot dz.\label{transitionary form}
\end{eqnarray}
In order to investigate the solutions for some
specific values of para\-meters appearing in the field functions that are
relevant to other models,
it is preferable  to redefine  these parameters as
\[
d=\frac{1}{2}(p+k-1),\quad q=d(d-(k-1)+\frac{1}{2}w(k-1)^2+k(k-1).
\]
These redefinitions bring the metric (\ref{transitionary form}) into the form
\begin{eqnarray}\label{metz}
g
=\left(1+c^2
r^{2d+1-k}\right)^2
\bigg\{r^{2d(d-k)+(\omega(k-1)+2k)(k-1)}(dr\ot dr-dt\ot dt)\nonumber\\
+W_0^2r^{2(k-d)}d\phi\ot d\phi \bigg\}
+
\left\{\frac{r^{d}}{1+c^2
r^{2d+1-k}}\right\}^2
dz\ot dz,\label{fin-form-of-g}
\end{eqnarray}
whereas the magnetic field strength $f$ becomes
\begin{equation}\label{fz}
f=\pm\sqrt{2}c(2d+1-k)\frac{r^{d(k+1-d)-k^2-\omega/2\,(k-1)^2}}{(1+c^2r^{1+2d-k})^{2}}.
\end{equation}

In the  the form given in (\ref{fin-form-of-g}), it is the easy to
probe the limiting cases of the metric. For
$k=1$ this solution reduces to corresponding Einstein-Maxwell
solution \cite{SKMHH,Bonnor}. For $c=0$ the magnetic field
vanishes and one obtains the static Brans-Dicke-Maxwell vacuum generalization of the
Levi-Civita vacuum solution  {\cite{Arazi}}:
\begin{eqnarray}\label{bdlc}
g
&=&
r^{2d(d-k)+(\omega(k-1)+2k)(k-1)}(dr\ot dr-dt\ot dt)+r^{2d}dz\ot dz
\nonumber\\
&&+
W_0^2r^{2(k-d)}d\phi\ot d\phi,\\
\phi&=&r^{1-k}.
\end{eqnarray}
Furthermore, for  $c=0$ and  $k=1$,
one  obtains the vacuum Levi-Civita solution of the form (\ref{metric1}):
\begin{equation}
g=r^{2d(d-1)}(-dt\ot dt+dr\ot dr )+r^{2d}dz\ot dz +W_0^2 r^{-2(d-1)}d\phi\ot d\phi.
\end{equation}
The solution (\ref{metz})-(\ref{fz}) have five constant parameters $d,W_0,k,\omega,c$.
The first two parameters are the parameters of the Levi-Civita solution which we will discuss below.
The next two parameters are
related with  the nontrivial coupling of  Brans-Dicke scalar to the metric.
The last parameter represents Maxwell fields.
Compared to Newtonian gravity generated by a line source which contains a single parameter related
to linear mass density, in general theory of relativity, there appear  two parameters.
For this metric, the  parameter $d$ is related to the energy
density of the source  of the  metric and the second parameter $W_0$ is related to the
global topology (conicity) of space-time \cite{Bonnorlc}. In
order to understand the meaning and the behavior of these
parameters, several interior cylinders \cite{Bonnorlc,cylinder} and
thin shells \cite{shells} have been constructed as sources of this
vacuum solution. For $d=0$ this metric describes a straight cosmic
string \cite{cosmicstring}. Thus, this solution can be considered
to describe a magnetic line source or a magnetic string in
Brans-Dicke-Maxwell theory.

\subsection{A Magnetic Field Along the $\phi$-Direction.}
For a magnetic field lines circulating  along the $\phi$-directions, one  has
$F=f\theta^{1}\wdg \theta^{3}$. This magnetic field  is apparently due to the 4-current
form along the $z$-axis. This solution can be found  either by solving the
field equations or by exchanging the coordinates $\phi$ and $z$.
The Maxwell equations are satisfied for $f=\beta e^{-K}$ with
$\beta$ being a constant. Skipping  the details of straightforward computations,
solutions to the field equations assume the form
\begin{eqnarray}
g
&=&
(1+c^2r^{1+k-2d})^2\bigg\{r^{2d(d-k)+(\omega(k-1)+2k)(k-1)}(dr\ot dr-dt\ot dt)
\nonumber\\
&&+r^{2d}dz\ot dz
\bigg\}
+
\left\{\frac{W_0r^{(k-d)}}{1+c^2r^{1+k-2d}}\right\}^2
d\phi\ot d\phi,\\
\phi
&=&
W_0^{-1}r^{1-k}, \\
f
&=&
\pm c\sqrt{2}(2d-k-1)
\frac{r^{-d(d-k+1)-(k+w/2(k-1))(k-1)}}{(1+c^2r^{1+k-2d})^{2}}\,.
\end{eqnarray}
As in the previous case, for the same values of the
parameters, one has the same special cases for this solution as well. Also for
$d=0$, the solution reduces to the Brans-Dicke version of the
Bonnor-Melvin universe \cite{BonnorMelvin} described in terms of the metric
\begin{eqnarray}
g=(1+c^2r^{1+k})^2\bigg\{r^{(\omega(k-1)+2k)(k-1)}(dr\ot dr-dt\ot dt)
+dz\ot dz
\bigg\}\nonumber\\
+
\left\{\frac{W_0r^{k}}{1+c^2r^{1+k}}\right\}^2d\phi\ot d\phi.\label{BonnorMelvin}
\end{eqnarray}
For $k=1$, (\ref{BonnorMelvin}) reduces to Einstein-Maxwell solution with the metric given by
\begin{eqnarray}
g=(1+c^2r^{2})^2\bigg\{-dt\ot dt+dr\ot dr +dz\ot dz
\bigg\}+\left\{\frac{W_0r}{1+c^2r^{2}}\right\}^2d\phi\ot d\phi.
\end{eqnarray}
Note that the Bonnor-Melvin solution is not singular at the axis
but the corresponding Brans-Dicke version is singular. The singularity vanishes
when $k=1$, however for this value of $k$, the Brans-Dicke theory reduces
to Einstein's theory.

\subsection{A Radial Electric Field}
The solution is easily found when one considers the metric
(\ref{metric2}). In this case, one starts with the closed Faraday 2-form $F=f(r)
\tilde{\theta}^{0}\wedge \tilde{\theta}^{1}$. Maxwell's equation $d*F=0$
yield $f(r)=e^{2U-K}/W$ (up to a constant).
Going through similar computations as in the preceding cases, one concludes
that the metric, scalar field and $F_{01}$ depend on the radial coordinate as
\begin{eqnarray}
g
&=&
-\left\{\frac{r^{2\sigma}}{1-c^2r^{4\sigma+1-k}}\right\}^2dt\ot dt+
(1-c^2r^{4\sigma+1-k})^2\times  \\
&&\bigg\{r^{4\sigma(2\sigma-k)+(w(1-k)-2k)(1-k)}\big(dr\ot dr+dz\ot dz\big)+ W_0^2
r^{2(k-\sigma)}d\phi\ot d\phi \bigg\}, \nonumber
\end{eqnarray}
\begin{eqnarray}
\phi
&=&
\frac{r^{(1-k)}}{W_0}, \\
f
&=&
\pm \sqrt{2}c(4\sigma+1-k)
\frac{r^{-2\sigma(2\sigma-1)-w/2(1-k)^2+k(2(1-\sigma)-k)}}{(1-c^2r^{4\sigma+1-k})^{2}}.
\end{eqnarray}
This solution can also be found by replacing $t\rightarrow i z$,
$z\rightarrow i t$, $c\rightarrow i c$, $d\rightarrow 2 \sigma$
from the solution corresponding to a magnetic field in $z$
direction. The solution is singular at $r=c^{-2/(4\sigma+1-k)}$
and this could be taken as the boundary of the source.
This solution reduces to Brans-Dicke vacuum solution when $c=0$,
 the Einstein-Maxwell solution with radial electric field when $k=1$.
 If $k=1$ and $c=0$, then the metric  reduces to the conventional form of
the static Levi-Civita solution:
\begin{equation}
g
=
-r^{4\sigma}dt\ot dt+ r^{4\sigma(2\sigma-1)}\big(dr\ot dr+dz\ot dz\big)+ W_0^2
r^{2(1-\sigma)}d\phi\ot d\phi.
\end{equation}
For $\sigma=0$, the Levi-Civita solution reduces to the Minkowski
metric and for Einstein-Maxwell  solution, the electric field also
vanishes, as one expects on physical grounds. However, for the
Brans-Dicke-Maxwell vacuum solution, when $\sigma=0$, the electric
field does not vanish. Hence, the solution
\begin{eqnarray}
g
&=&
-\frac{dt\ot dt}{(1-c^2r^{1-k})^2}\,+
(1-c^2r^{1-k})^2\times  \\
&&\bigg\{r^{(-2k+\omega(1-k))(1-k)}\big(dr\ot dr+dz\ot dz\big)+ W_0^2
r^{2k}d\phi\ot d\phi\bigg\}, \nonumber\\
\phi
&=&
W_0^{-1}r^{1-k},\\
f
&=&
\pm \sqrt{2}c(1-k)
\frac{r^{-\omega/2(1-k)^2+k(2-k)}}{(1-c^2r^{1-k})^{2}},\label{share}
\end{eqnarray}
is only present for Brans-Dicke theory and the electric field
vanishes for $k=1$. Similar argument holds for the solution with a
magnetic field in $z$-direction.

Finally, we note that it is possible to start with
$F=f(r)\tilde\theta^{0}\wdg\tilde\theta^{1}+h(r)\tilde\theta^{2}\wdg\tilde\theta^{3}$,
then one ends up with the same metric and the scalar function, for which the
diagonal energy momentum
form corresponding to (\ref{share}) is shared between a magnetic and an electric field.

\section{Discussion}
This work presents exact solutions of the Brans-Dicke-
Maxwell theory in static cylindrical symmetry. The solutions
reduce to corresponding Einstein-Maxwell solutions when a constant
set to a specific value ($k=1$), to the Brans-Dicke vacuum solution
when a constant related with field strength vanishes ($c=0$), or to the
general relativistic vacuum Levi-Civita solution when both the
constants are set to these specific values. Unlike the
Einstein-Maxwell solutions, there is no solution which is
everywhere regular in the Brans-Dicke-Maxwell case.

It is  interesting to note that, for all the cases considered above,  although the
scalar $\phi$  does not depend on
the Brans-Dicke parameter $\omega$, the field strengths $F$
do have $\omega$ dependence.
Therefore, the general relativistic limit of the above theory can be
recovered by simply setting the metric parameter $k=1$, (which entails $1/\phi=W_0$)
rather then by inspecting  the field equations in the $\omega\mapsto\infty$ limit
of the Brans-Dicke parameter.
Understanding  the significance and implications of  the parameter
$\omega$ and its physically plausible numerical
values in the model spacetimes presented above, needs further study. For example, the
investigation of the motion of charged test particles in spacetime models constructed above in Brans-Dicke theory.

\section*{Acknowledgements}
We thank Prof. Metin Ar\i k for reading the manuscript and helpful discussions.

\end{document}